\documentclass[final,5p,times,twocolumn,numbers]{elsarticle}

\usepackage{amssymb}
\usepackage{amsmath}
\usepackage{lipsum}
\usepackage{enumitem}

\usepackage{bm}
\usepackage{algorithm}
\usepackage{algpseudocode}

\begin{document}

\begin{frontmatter}



\title{Adapting Quantum Machine Learning for Energy Dissociation of Bonds}

\author[1]{Swathi Chandrasekhar}
\author[1]{Shiva Raj Pokhrel}
\author[1]{Navneet Singh}

\affiliation[1]{organization={School of IT, Deakin University},
            city={Geelong},
            postcode={3122}, 
            state={Victoria},
            country={Australia}}

\begin{abstract}
Accurate prediction of bond dissociation energies (BDEs) underpins mechanistic insight and the rational design of molecules and
materials. We present a systematic, reproducible benchmark comparing quantum and classical machine learning models for BDE
prediction using a chemically curated feature set encompassing atomic properties (atomic numbers, hybridization), bond characteristics (bond order, type), and local environmental descriptors. Our quantum framework, implemented in Qiskit Aer on six qubits,
employs ZZFeatureMap encodings with variational ans¨ atz (RealAmplitudes) across multiple architectures Variational Quantum Regressors (VQR), Quantum Support Vector Regressors (QSVR), Quantum Neural Networks (QNN), Quantum Convolutional Neural
Networks (QCNN), and Quantum Random Forests (QRF). These are rigorously benchmarked against strong classical baselines,
including Support Vector Regression (SVR), Random Forests (RF), and Multi-Layer Perceptrons (MLP). Comprehensive evalua-
tion spanning absolute and relative error metrics, threshold accuracies, and error distributions shows that top-performing quantum
models (QCNN, QRF) match the predictive accuracy and robustness of classical ensembles and deep networks, particularly within
the chemically prevalent mid-range BDE regime. These findings establish a transparent baseline for quantum-enhanced molecular property prediction and outline a practical foundation for advancing quantum computational chemistry toward near chemical
accuracy.

\end{abstract}




\end{frontmatter}




\section{Introduction}

Roy J. Plunkett's discovery~\cite{plunkett1986history} of polytetrafluoroethylene (PTFE, commercialized as Teflon) during tetrafluoroethylene experiments revolutionized materials science. PTFE's extraordinary chemical inertness and thermal stability stem directly from exceptionally strong carbon–fluorine bonds, which demand several hundred kilojoules per mole to break \cite{nayak2024computing}. These formidable bond dissociation energies (BDEs) underpin PTFE's resistance to solvents, acids, bases, and extreme conditions, enabling its widespread industrial applications. Yet, PTFE and related fluoropolymers also exemplify the environmental persistence of high-BDE materials, raising global concerns over ``forever chemicals'' and their ecological impact. This dual legacy underscores a critical principle: the bond dissociation energy serves as the cornerstone property governing both molecular utility and long-term environmental fate \cite{liu2025atmospheric, gao2025environmental}.

BDE values are indispensable across chemistry and materials science, driving mechanistic insights, enabling rational molecular design, and informing synthesis safety and efficiency. However, they also dictate degradation pathways, pollutant persistence, and the environmental half-lives of synthetic compounds. Traditional BDE determination methods face severe limitations: experimental measurements are prohibitively time-consuming and resource-intensive, while high-accuracy quantum chemical calculations (e.g., coupled cluster or composite methods) become computationally intractable for larger molecular systems \cite{cao2019quantum}. This computational bottleneck has catalyzed the development of data-driven surrogate models for rapid BDE prediction, crucial not only for advancing molecular design but also for anticipating and mitigating environmental risks \cite{sun2023combined}.

Classical machine learning \cite{bin2025exploring} approaches have demonstrated remarkable success in BDE prediction by extracting structure–property relationships from curated datasets. These surrogate models, trained on quantum chemical data, deliver predictions in milliseconds rather than days, approximating computationally intensive electronic structure calculations with impressive accuracy \cite{liu2024prediction}. This acceleration enables the exploration of vast chemical spaces entirely inaccessible to traditional \textit{ab initio} methods, supporting the discovery of sustainable catalysts, degradable materials, and safer chemical substitutes. However, classical approaches increasingly struggle with representation learning, scaling challenges, and accuracy maintenance as datasets grow in size and complexity.

Quantum machine learning (QML)~\cite{huang2020quantum} offers a promising paradigm for environmentally conscious chemistry. By leveraging quantum mechanical principles directly, quantum feature maps and models can potentially capture complex, high-dimensional structures that remain elusive to classical kernels and featurization techniques~\cite{kumar2025quantum}. This intrinsic quantum advantage may prove particularly valuable for modelling molecular properties fundamentally governed by quantum effects, such as bond strength, stability, and degradation pathways.  In this work, we present a head-to-head comparison between classical and quantum pipelines for BDE prediction using the comprehensive BDE-db dataset \cite{st2020prediction} with advanced feature engineering. We deliver a reproducible benchmark to identify scenarios where QML demonstrates tangible advantages for chemical property prediction. By doing so, we establish a practical foundation for quantum-enhanced molecular design tools that can accelerate the discovery of environmentally sustainable materials and support global efforts to reduce chemical pollution \cite{jin2024insight}.

\subsection{Literature Review}
\begin{table*}[t]
\centering
\caption{Comparative Analysis of ML Approaches for Chemical Property Prediction}
\label{tab:ml_comparison}
\begin{tabular}{|p{1.348cm}|p{1.892cm}|p{3.5cm}|p{3.5cm}|p{3.5cm}|p{2.1cm}|}
\hline
\textbf{Approach} & \textbf{Methods} & \textbf{Ideas} & \textbf{Limitations} & \textbf{BDE Performance} & \textbf{Refs.} \\
\hline
\textbf{ML} & Random Forest, SVR, Neural Networks & Established methodology; Fast inference; Extensive validation & Fixed feature representations; Limited quantum effects modeling & Near-chemical accuracy ($\sim$3-5 kcal/mol); Sub-second prediction & \cite{st2020prediction,nayak2024computing} \\
\hline
\textbf{QML} & QSVM, QNN, VQR, QCNN & Natural representation of quantum effects; Potential exponential speedup; Handles high-dimensional feature spaces & Hardware limitations; Optimization challenges; Limited chemical applications & Emerging capability; Comparable to classical ML in limited studies & \cite{sajjan2022quantum,huang2020quantum}~\cite{patel2025quantum, kumar2025quantum,ajagekar2020quantum} \\
\hline
\textbf{Hybrid} & Quantum kernels with classical optimization & Combines quantum feature mapping with classical robustness & Requires quantum-classical interface; Communication overhead & Promising for specific molecular properties; Active research area & \cite{sajjan2022quantum,patel2025quantum, ajagekar2020quantum} \\
\hline
\textbf{Our work} & SVR/QSVR, RF/QRF, MLP/QNN & Direct comparison on identical dataset; Rigorous error analysis & Limited to simulated quantum hardware & Quantum models approach classical performance; QRF and QCNN most promising & -- \\
\hline
\end{tabular}
\end{table*}
ML has transformed chemical research by uncovering patterns in complex datasets and automating predictive tasks that were once labor-intensive. Over the past two decades, advances in large-scale chemical databases, computational resources, and algorithmic innovation have enabled efficient dimensionality reduction and data manipulation, driving breakthroughs across chemistry \cite{sajjan2022quantum}. Today, ML underpins progress in electronic structure prediction, materials discovery, retrosynthetic planning, and reaction control, with particularly strong impact in physical chemistry and chemical physics. A detailed comparison of these developments is shown in Tables~1 and 2.

John et al.~\cite{st2020prediction} demonstrated that classical ML models can predict BDEs for organic molecules with near-chemical accuracy at sub-second speeds. Their automated density functional theory (DFT) workflow on more than 42,000 molecules generated a dataset of nearly 300,000 BDEs, providing a foundation for surrogate modeling. Other neural network approaches have attempted to predict bond contributions to atomization energies, though with higher errors ($\sim$10 kcal/mol) relative to experimental benchmarks \cite{nayak2024computing}.

Despite these advances, classical ML approaches face persistent limitations: reliance on fixed molecular descriptors, challenges in representation learning, high computational costs for generating training datasets, limited interpretability of ``black box’’ models, and poor transferability to larger or more diverse molecular systems \cite{st2020prediction,huang2020quantum,patel2025quantum}.

QML has emerged as a promising alternative, leveraging quantum principles such as superposition, entanglement, and interference to potentially deliver computational advantages for molecular property prediction \cite{sajjan2022quantum}. QML algorithms are particularly well-suited for modeling quantum many-body effects that govern bond breaking and formation, making them attractive for accurate BDE prediction. Although applications of QML to BDE prediction remain at an early stage, the potential is significant: quantum models could enable rapid and scalable prediction across diverse molecular spaces, supporting high-throughput screening and sustainable compound discovery.

In this work, we address a key gap in the literature by presenting a systematic benchmark of classical and quantum ML approaches for BDE prediction. We evaluate classical baselines (SVR, Random Forest, MLP) alongside quantum counterparts (QSVR, QRF, QNN), providing a reproducible comparison on a real-world chemical property prediction task.

\begin{table}[ht]
\centering
\footnotesize
\caption{Classical vs. Quantum ML Methods for Chemical Applications}
\label{tab:ml_qml_summary}
\begin{tabular}{|p{1.025cm}|p{1.62cm}|p{2.35cm}|p{2.1cm}|}
\hline
\textbf{Algos} & \textbf{Application} & \textbf{Performance} & \textbf{Limitations} \\
\hline
QSVM & Classification, Regression & Potential exponential speedup; few chemical benchmarks & Limited chemical applications  \\
\hline
QNN & Feature Extraction, Phase Classification & Applied to quantum/classical data & Mostly hybrid, interpretability  \\
\hline
VQR & State Preparation, Tomography & Reduces quantum resource needs & Barren plateaus, optimization \\
\hline
QCNN & Phase Recognition & Hierarchical feature extraction & Early stage, scalability \\
\hline
CNN, SVM, RF & BDE prediction, property modeling & Near-chemical accuracy, fast prediction & Data quality, generalization, interpretability \\
\hline
\end{tabular}
\end{table}

\subsection{Key Contributions}

\begin{enumerate}[label=\alph*)]

    \item \textit{Unified benchmark:}  
    We release a reproducible pipeline for BDE prediction spanning data curation, feature engineering, model training, and evaluation across classical and quantum ML (Qiskit Aer, scikit-learn).

    \item \textit{Rigorous dataset curation:}  
    Chemically validated BDE dataset with SMILES integrity checks, bond index verification, outlier removal (5--7\%), and stratified sampling across bond classes.

    \item \textit{Comprehensive QML suite:}  
    Implementation of VQR, QSVR, QNN, QCNN, and QRF to probe diverse inductive biases.

    \item \textit{Classical vs. quantum comparison:}  
    Benchmarking against SVR, RF, and MLP with standardized preprocessing and cross-validated optimization.

    \item \textit{Performance in mid-range BDEs:}  
    QCNN and QRF achieve accuracy comparable to RF and MLP in the 70--100 kcal/mol regime relevant to practical chemistry.

    \item \textit{Error structure analysis:}  
    All models exhibit a U-shaped error profile, with lowest errors in mid-range BDEs and higher errors at extremes ($<$70, $>$100 kcal/mol).

    \item \textit{Roadmap to near-chemical accuracy:}  
    Strategy combining high-fidelity labels, enriched descriptors, multi-fidelity $\Delta$-learning, and calibrated ensembles, positioning QML as a complementary tool for accuracy, robustness, and interpretability.

\end{enumerate}

\section{Methodology and Implementation}

Figure~1 shows the proposed methodological pipeline for the BDE prediction: (a) curation and validation of the BDE-db dataset, (b) feature engineering and scaling to encode atomic, bond, and environmental descriptors, (c) supervised training of quantum and classical machine learning models, and (d) predictive inference and evaluation. The curated dataset preserves chemical diversity across common and rare bond types, while engineered features capture atomic properties, bond characteristics, and local environments, providing a robust foundation for model benchmarking. We discuss the details of each of them in the following.

\begin{figure*}[h]
  \centering
  \includegraphics[width = 1.79\columnwidth]{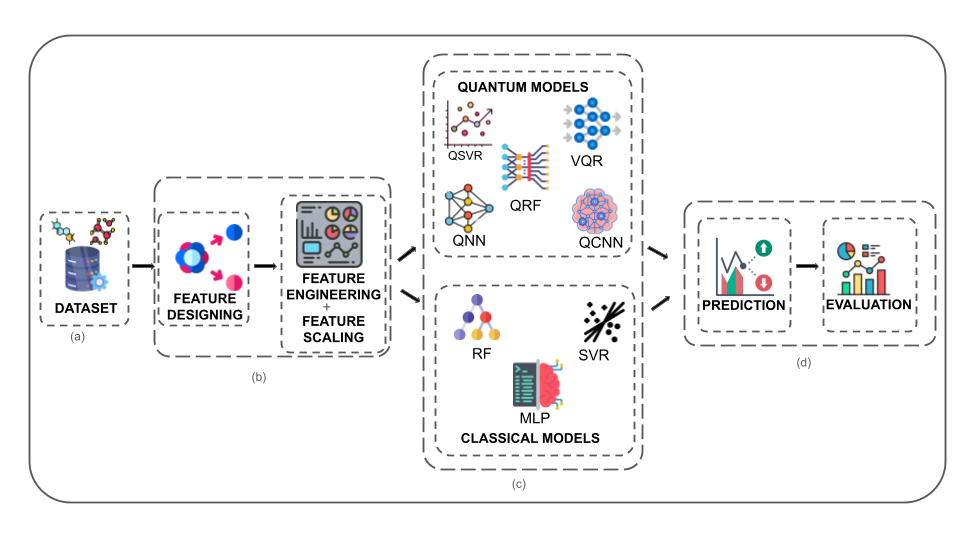}
  \caption{(a) BDE-db Dataset, (b) feature designing, engineering, and scaling to capture atomic, bond, and environmental descriptors relevant to dissociation energetics, (c) Traning of Quantum and classical models, (d) Prediction and evaluation of the model}
  \label{fig:architecture}
\end{figure*}
\subsection{Dataset Preparation}

The BDE-db dataset~\cite{st2020prediction} provides a comprehensive collection of molecular structures, bond indices, and experimentally measured BDEs . To ensure chemical validity and reliability, the dataset underwent a rigorous preprocessing workflow. This included verification of SMILES representations, validation of bond indices, and systematic outlier detection. SMILES (Simplified Molecular Input Line Entry System) encodes molecular structures as text strings and serves as the primary representation format. Through this validation pipeline, approximately 5--7\% of erroneous or inconsistent entries were identified and removed, thereby ensuring that the final dataset consisted exclusively of chemically meaningful and structurally valid molecules suitable for robust model training.  

To achieve a balance between computational tractability and chemical diversity, we employed a stratified sampling approach. This strategy preserved representation across all major bond classes, including C--H, C--C, and C--O, while simultaneously ensuring sufficient coverage of rarer bond types. In doing so, the natural statistical distribution of common bonds was retained, yet rare categories were not underrepresented. The outcome of this sampling design was a chemically diverse and computationally efficient dataset, thereby providing an adequate basis for model development and comparative analysis.  

Feature engineering was undertaken to encode the physicochemical descriptors most relevant to bond dissociation energetics. For each bond, atomic-level descriptors such as atomic numbers and hybridization states of the participating atoms were extracted, capturing the electronic structures of the radical fragments produced upon dissociation. Bond order was explicitly incorporated, as it fundamentally determines bond strength and governs chemical reactivity. The local chemical environment surrounding each bonded atom was quantified in terms of neighboring heteroatoms (C, O, N), thus encoding inductive and resonance effects that modulate bond energetics. Furthermore, categorical encoding was applied to bond types to ensure compatibility with machine learning algorithms while retaining the chemical identity of each bond.  

The resulting feature representation integrates both quantum-mechanical and classical determinants of bond stability. This carefully curated dataset provides a chemically rigorous and computationally tractable foundation for benchmarking predictive models, enabling a direct and systematic comparison between quantum algorithms and traditional machine learning approaches for molecular bond energetics.

\subsection{Classical ML Models}

Support Vector Regression (SVR)  
is implemented with a radial basis function (RBF) kernel, owing to its capacity to capture nonlinear relationships in high-dimensional feature spaces. SVR estimates a regression function $f(\mathbf{x})$ that deviates from the true target $y$ by at most a tolerance $\epsilon$, while penalizing deviations beyond this margin through slack variables\cite{zhang2020support}. The optimization problem is formulated as
\[
\min_{\mathbf{w}, b} \ \frac{1}{2} \|\mathbf{w}\|^2 + C \sum_{i=1}^N (\xi_i + \xi_i^*),
\]
subject to

\begin{equation*}
\begin{aligned}
y_i - \langle \mathbf{w}, \phi(\mathbf{x}_i) \rangle - b &\leq \epsilon + \xi_i, 
\langle \mathbf{w}, \phi(\mathbf{x}_i) \rangle + b - y_i &\leq \epsilon + \xi_i^*, \\
\xi_i, \xi_i^* &\geq 0.
\end{aligned}
\end{equation*}

where $C$ is the regularization parameter, $\phi(\cdot)$ denotes the nonlinear feature mapping, and $\xi_i, \xi_i^*$ are slack variables that permit violations of the $\epsilon$-insensitive margin. Hyperparameters $C$, $\gamma$, and $\epsilon$ were systematically optimized via grid search with cross-validation.  

With Random Forest (RF)  as 
 a nonparametric ensemble method, we anticipate overfitting by averaging predictions across multiple randomized decision trees\cite{segal2004machine}. Each tree is trained on a bootstrap-resampled dataset, and at each split, a random subset of features is selected to maximize information gain. The ensemble prediction is given by
\[
\hat{y} = \frac{1}{T} \sum_{t=1}^T f_t(\mathbf{x}),
\]
where $T$ is the number of trees and $f_t(\mathbf{x})$ denotes the prediction of the $t$-th tree. The key hyperparameters including the number of trees, maximum tree depth, and minimum number of samples per split were optimized through grid search with cross-validation to ensure model robustness.  

Further, Multi-Layer Perceptron (MLP) 
 is implemented as a fully connected feedforward neural network capable of learning highly nonlinear mappings between inputs and targets\cite{kruse2022multi}. Given an input $\mathbf{x}_i$, the MLP computes a nonlinear transformation $f(\mathbf{x}_i; \mathbf{W}, \mathbf{b})$ parameterized by weights $\mathbf{W}$ and biases $\mathbf{b}$. The network is trained by minimizing the mean squared error (MSE):  
\[
\min_{\mathbf{W}, \mathbf{b}} \ \frac{1}{N} \sum_{i=1}^N \left( y_i - f(\mathbf{x}_i; \mathbf{W}, \mathbf{b}) \right)^2.
\]
Nonlinear activation functions introduce expressivity, enabling the network to approximate complex functions. Hyperparameters, including the number of hidden layers, number of units per layer, activation functions, and regularization strength, were optimized using cross-validated grid search to prevent overfitting while maximizing predictive accuracy.

It should be noted that all classical ML models were trained on standardized and normalized feature sets to ensure numerical stability and comparability between algorithms. Hyperparameter optimization was performed using nested cross-validation to mitigate selection bias. Model performance was evaluated using multiple metrics, including MSE, mean absolute error (MAE), and the coefficient of determination ($R^2$), providing a comprehensive assessment of predictive accuracy and generalization capability.

\subsection{Quantum Machine Learning Models}

Classical molecular features were first normalized and encoded into quantum states via a parameterized feature map, specifically the ZZFeatureMap \cite{singh2025modeling, singh2025modelinggenomic}. For each input vector $\vec{x} \in \mathbb{R}^d$, the feature map applies data-driven rotations across $d$ qubits, producing the quantum state $|\phi(\vec{x})\rangle$. This encoding enables the circuit to embed complex, high-dimensional classical data into the Hilbert space, thereby leveraging quantum superposition and entanglement to enhance expressivity \cite{torlai2020machine}.  

The ZZFeatureMap implements angle encoding as
\[
\Phi(\mathbf{x}) = \bigotimes_{i=1}^{n} R_Z(x_i) \bigotimes_{(i,j) \in E} R_{ZZ}(\theta_{ij}),
\]
where $R_Z(x_i) = e^{-i\frac{x_i}{2}Z}$ denotes a rotation about the $Z$-axis and $R_{ZZ}(\theta_{ij}) = e^{-i\frac{\theta_{ij}}{2}Z_i \otimes Z_j}$ represents the entangling interaction between qubits $i$ and $j$.  

A variational ansatz was constructed using the RealAmplitudes circuit to introduce trainable parameters and enhance circuit flexibility:
\[
U(\bm{\theta}) = \prod_{l=1}^{L} \left( \bigotimes_{i=1}^{n} R_Y(\theta_{l,i}) \bigotimes_{(i,j) \in E} R_Z(\theta_{l,ij}) \right),
\]
where $L$ is the number of layers and $\bm{\theta}$ are the trainable parameters. The resulting parameterized quantum state is therefore
\[
|\psi(\mathbf{x}; \bm{\theta})\rangle = U(\bm{\theta}) \Phi(\mathbf{x}) |\mathbf{0}\rangle,
\]
which serves as the input for the variational quantum regressor.

All quantum circuits were implemented using Qiskit and executed on the Aer simulator to ensure reproducibility without specialized hardware. The circuits employed six qubits with ten repetitions of the feature map and ten layers of the variational ansatz, using linear entanglement connectivity. The target variable was normalized to the range $[-1,1]$ to improve convergence of the quantum training process. Optimization of circuit parameters was performed using the COBYLA optimizer with a maximum of 1000 iterations.

\begin{algorithm}[ht]
\caption{Variational Quantum Regressor (VQR)}
\label{alg:vqr}
\begin{algorithmic}[1]
\Procedure{VQR\_Train}{$\{(\vec{x}_i, y_i)\}_{i=1}^N$, feature\_map, ansatz, observable, optimizer}
    \State Normalize and scale input features $\vec{x}_i$
    \State Initialize parameters $\vec{\theta}$ for ansatz
    \Repeat
        \For{each training sample $(\vec{x}_i, y_i)$}
            \State Prepare quantum state: $|\psi(\vec{x}_i; \vec{\theta})\rangle = U(\vec{\theta}) \Phi(\vec{x}_i) |\mathbf{0}\rangle$
            \State Predict: $\hat{y}_i = \langle \psi(\vec{x}_i; \vec{\theta}) | \hat{O} | \psi(\vec{x}_i; \vec{\theta}) \rangle$
        \EndFor
        \State Compute loss: $\mathcal{L} = \frac{1}{N} \sum_{i=1}^N (y_i - \hat{y}_i)^2$
        \State Update $\vec{\theta}$ using optimizer
    \Until{convergence}
    \State \textbf{return} trained parameters $\vec{\theta}$
\EndProcedure
\end{algorithmic}
\end{algorithm}

\subsubsection{\textbf{Variational Quantum Regressor (VQR):}}
We implement  VQR
as a hybrid quantum-classical model designed to predict bond dissociation energies by learning a parameterized quantum circuit. See Algorithm~1 for details. For each input $\vec{x}$, the circuit prepares the state $|\psi(\vec{x}; \bm{\theta})\rangle = U(\bm{\theta}) \Phi(\vec{x}) |\mathbf{0}\rangle$. Predictions are obtained as the expectation value of a chosen observable $\hat{O}$:
\[
\hat{y}(\vec{x}) = \langle \psi(\vec{x}; \bm{\theta}) | \hat{O} | \psi(\vec{x}; \bm{\theta}) \rangle.
\]
The trainable parameters $\bm{\theta}$ are optimized by minimizing the mean squared error between predicted and true BDE values:
\[
\mathcal{L}(\bm{\theta}) = \frac{1}{N} \sum_{i=1}^N \left( y_i - \hat{y}(\vec{x}_i) \right)^2.
\]
Optimization proceeds iteratively, alternating between quantum circuit evaluations and classical parameter updates via COBYLA, thereby forming a fully hybrid quantum-classical training loop \cite{wang2023variational}.

\begin{algorithm}[ht]
\caption{Quantum Support Vector Regressor (QSVR)}
\label{alg:qsvr}
\begin{algorithmic}[1]
\Procedure{QSVR\_Train}{$\{(\vec{x}_i, y_i)\}_{i=1}^N$, feature\_map, SVR\_params}
    \State Normalize and scale input features $\vec{x}_i$
    \For{each pair $(\vec{x}_i, \vec{x}_j)$}
        \State Prepare quantum states $|\phi(\vec{x}_i)\rangle$, $|\phi(\vec{x}_j)\rangle$ using feature\_map
        \State Compute kernel: $K_{ij} = |\langle \phi(\vec{x}_i) | \phi(\vec{x}_j) \rangle|^2$
    \EndFor
    \State Train SVR using quantum kernel matrix $K$ and targets $y_i$
    \State \textbf{return} trained SVR model
\EndProcedure
\end{algorithmic}
\end{algorithm}

\subsubsection{\textbf{Quantum Support Vector Regressor (QSVR):}}
QSVR extends classical SVR by employing a quantum kernel, which implicitly maps data into a quantum Hilbert space \cite{suzuki2024quantum}. The kernel is defined as:
\[
K(\vec{x}_i, \vec{x}_j) = |\langle \phi(\vec{x}_i) | \phi(\vec{x}_j) \rangle|^2
\]
where $|\phi(\vec{x})\rangle$ is the quantum state prepared by the feature map. The SVR dual optimization problem is solved using this kernel, enabling the model to capture complex, non-linear relationships in the data. See Algorithm~2 for details.

\begin{algorithm}[ht]
\caption{Quantum Neural Network (QNN) Regressor}
\label{alg:qnn}
\begin{algorithmic}[1]
\Procedure{QNN\_Train}{$\{(\vec{x}_i, y_i)\}_{i=1}^N$, feature\_map, ansatz, observable, optimizer}
    \State Normalize and scale input features $\vec{x}_i$
    \State Initialize network parameters $\vec{\theta}$
    \Repeat
        \For{each training sample $(\vec{x}_i, y_i)$}
            \State Prepare quantum state: $|\psi(\vec{x}_i; \vec{\theta})\rangle = U(\vec{\theta}) \Phi(\vec{x}_i) |\mathbf{0}\rangle$
            \State Predict: $\hat{y}_i = \langle \psi(\vec{x}_i; \vec{\theta}) | \hat{O} | \psi(\vec{x}_i; \vec{\theta}) \rangle$
        \EndFor
        \State Compute loss: $\mathcal{L} = \frac{1}{N} \sum_{i=1}^N (y_i - \hat{y}_i)^2$
        \State Update $\vec{\theta}$ using optimizer
    \Until{convergence}
    \State \textbf{return} trained parameters $\vec{\theta}$
\EndProcedure
\end{algorithmic}
\end{algorithm}
\subsubsection{\textbf{Quantum Neural Network (QNN):}}
The QNN model constructs a quantum neural network by composing the feature map and ansatz circuits. For each input, the quantum state is measured with respect to an observable $\hat{O}$, yielding the prediction:
\[
\hat{y}(\vec{x}) = \langle \psi(\vec{x}; \vec{\theta}) | \hat{O} | \psi(\vec{x}; \vec{\theta}) \rangle
\]
See Algorithm~3 for details. The network parameters $\vec{\theta}$ are trained to minimize the MSE loss, using a hybrid quantum-classical optimization loop \cite{du2021learnability}.

\begin{algorithm}[ht]
\caption{Quantum Random Forest (QRF)}
\label{alg:qrf}
\begin{algorithmic}[1]
\Procedure{QRF\_Train}{$\{(\vec{x}_i, y_i)\}_{i=1}^N$, T, feature\_map, ansatz, observable, optimizer}
    \For{$t = 1$ to $T$}
        \State Sample bootstrap dataset $\mathcal{D}_t$
        \State Train VQR model $f_t$ on $\mathcal{D}_t$ using \Call{VQR\_Train}{}
    \EndFor
    \State \textbf{return} ensemble $\{f_1, \ldots, f_T\}$
\EndProcedure
\Procedure{QRF\_Predict}{$\vec{x}$, $\{f_1, \ldots, f_T\}$}
    \State Predict: $\hat{y}_{\text{QRF}}(\vec{x}) = \frac{1}{T} \sum_{t=1}^T f_t(\vec{x})$
    \State \textbf{return} $\hat{y}_{\text{QRF}}(\vec{x})$
\EndProcedure
\end{algorithmic}
\end{algorithm}

\subsubsection{\textbf{Quantum Random Forest (QRF):}}
QRF is an ensemble method where each base learner is a quantum tree, i.e., a variational quantum circuit trained on a bootstrap sample. For $T$ quantum trees\cite{srikumar2024kernel}, the final prediction is the average of individual tree outputs:
\[
\hat{y}_{\text{QRF}}(\vec{x}) = \frac{1}{T} \sum_{t=1}^T \hat{y}_t(\vec{x})
\]
where each $\hat{y}_t(\vec{x})$ is obtained from a separately trained VQR model. See Algorithm~4 for details. Observe that we leverage quantum diversity and ensemble averaging to improve predictive performance and robustness.

\subsubsection{\textbf{Quantum Convolutional Neural Network (QCNN):}}
The QCNN adapts classical convolutional principles to quantum circuits, leveraging quantum feature maps, convolutional layers, and pooling operations. For an input feature vector $\mathbf{x}$, the process begins by encoding it into a quantum state using a feature map $\Phi(\mathbf{x})$, typically a ZZFeatureMap. A quantum convolutional layer $U_{\mathrm{conv}}(\boldsymbol{\theta}_c)$ applies local, parameterized entangling gates (e.g., $R_Y$ rotations and CNOT gates) to adjacent qubits, with parameters $\boldsymbol{\theta}_c$ shared across these operations. This mimics classical weight sharing and extracts local features. See Algorithm~5 for details. 

 The prediction $\hat{y}(\mathbf{x})$ is obtained by measuring the expectation value of a chosen observable $\hat{O}$ (e.g., Pauli-$Z$ operators):
\[
\hat{y}(\mathbf{x}; \boldsymbol{\theta}) = \langle \phi(\mathbf{x}) | U^\dagger(\boldsymbol{\theta}) \, \hat{O} \, U(\boldsymbol{\theta}) | \phi(\mathbf{x}) \rangle,
\]
where $U(\boldsymbol{\theta})$ represents the combined QCNN circuit with all trainable parameters $\boldsymbol{\theta} = \{\boldsymbol{\theta}_c, \boldsymbol{\theta}_p, \boldsymbol{\theta}_d\}$. The model is trained by minimizing the mean squared error (MSE) loss:
\[
\mathcal{L}(\boldsymbol{\theta}) = \frac{1}{N} \sum_{i=1}^N \big( y_i - \hat{y}(\mathbf{x}_i; \boldsymbol{\theta}) \big)^2,
\]
using a hybrid quantum-classical optimizer like COBYLA.

\begin{algorithm}[ht]
\caption{Quantum Convolutional Neural Network (QCNN) Regressor}
\label{alg:qcnn}
\begin{algorithmic}[1]
\Procedure{QCNN\_Train}{$\{(\vec{x}_i, y_i)\}_{i=1}^N$, feature\_map, QCNN\_layers, final\_layer, observable, optimizer}
    \State Encode classical features into quantum states: $|\phi(\vec{x}_i)\rangle = \Phi(\vec{x}_i)|0\rangle^{\otimes n}$
    \State Apply QCNN layers (convolution and pooling operations)
    \State Apply final dense variational layer $U_{\mathrm{dense}}$
    \Repeat
        \For{each sample $(\vec{x}_i, y_i)$}
            \State Prepare full QCNN circuit $U(\boldsymbol{\theta})$
            \State Predict $\hat{y}_i = \langle \psi(\vec{x}_i;\boldsymbol{\theta}) | \hat{O} | \psi(\vec{x}_i;\boldsymbol{\theta}) \rangle$
        \EndFor
        \State Compute loss $\mathcal{L}(\boldsymbol{\theta})$
        \State Update $\boldsymbol{\theta}$ using optimizer
    \Until{convergence}
    \State \Return trained parameters $\boldsymbol{\theta}$
\EndProcedure
\end{algorithmic}
\end{algorithm}

\section{Performance Evaluation \& Results}

All experiments were conducted on a standard macOS workstation with 8~GB RAM and Apple Silicon architecture. The proposed methodology is hardware-agnostic and does not require specialized quantum processors or GPU resources, ensuring reproducibility across conventional computing platforms. Implementation was performed in Python (version 3.9 or later) using Qiskit (version 1.4.2) for quantum circuit construction, simulation, and machine learning integration. Specifically, modules utilized include \texttt{qiskit}, \texttt{qiskit.circuit}, \texttt{qiskit.circuit.library}, visualization tools (\texttt{qiskit.visualization}), compiler functionalities (\texttt{qiskit.compiler}), primitives, and the Aer backend (\texttt{qiskit\_aer}) for high-fidelity simulation. Quantum-enhanced algorithms were implemented using the Qiskit Machine Learning package, while classical data processing leveraged \textit{pandas} and \textit{NumPy}, with machine learning baselines developed using \textit{scikit-learn}.

\begin{figure}[h]
  \centering
  \includegraphics[width = \columnwidth]{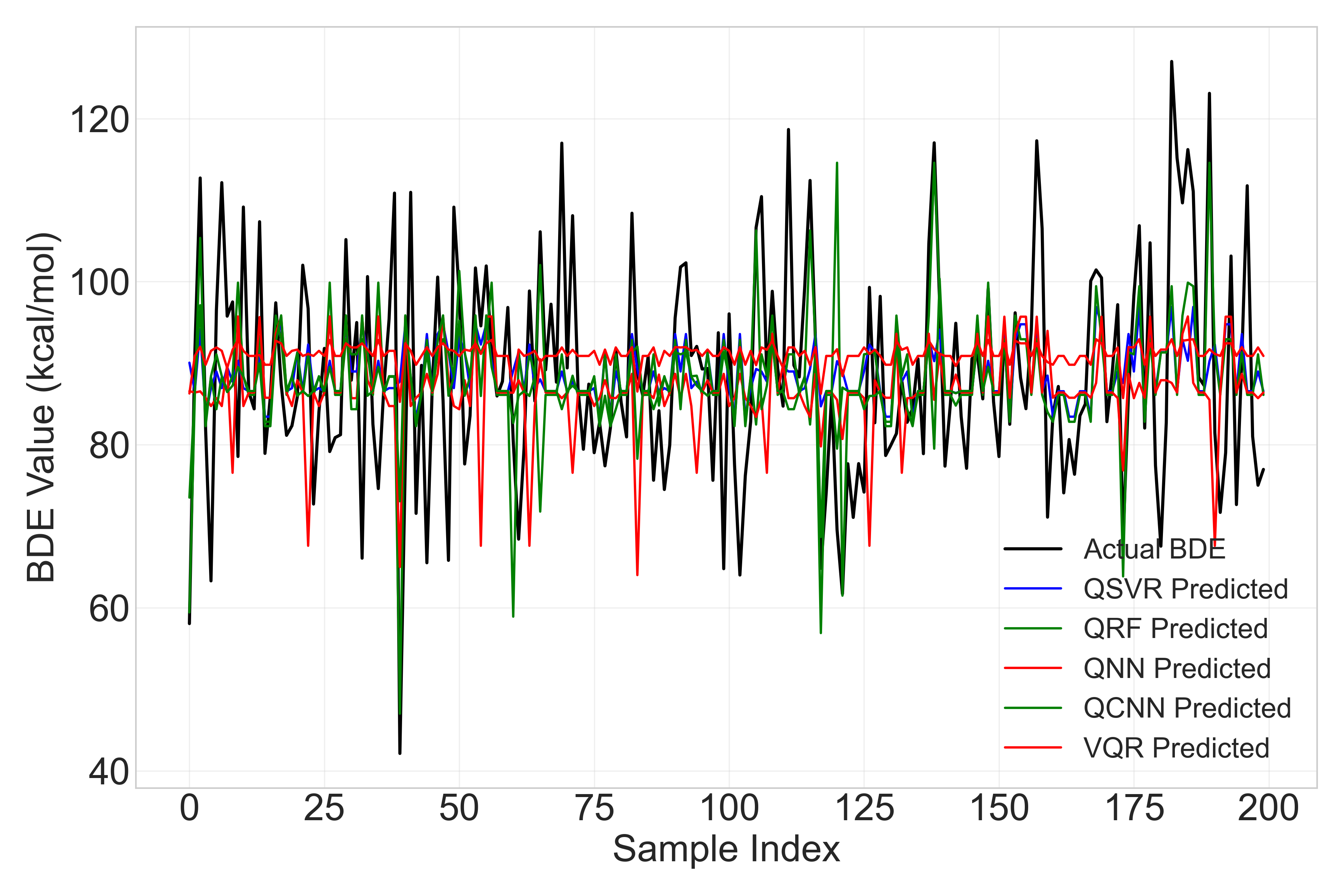}
  \caption{Comparison of Actual vs. Predicted BDE Values Across five quantum models: VQR, QSVR, QRF, QCNN, and QNN.}
  \label{fig:ActualVSPredictedBDE}
\end{figure}

\begin{figure}[h]
  \centering
  \includegraphics[width = \columnwidth]{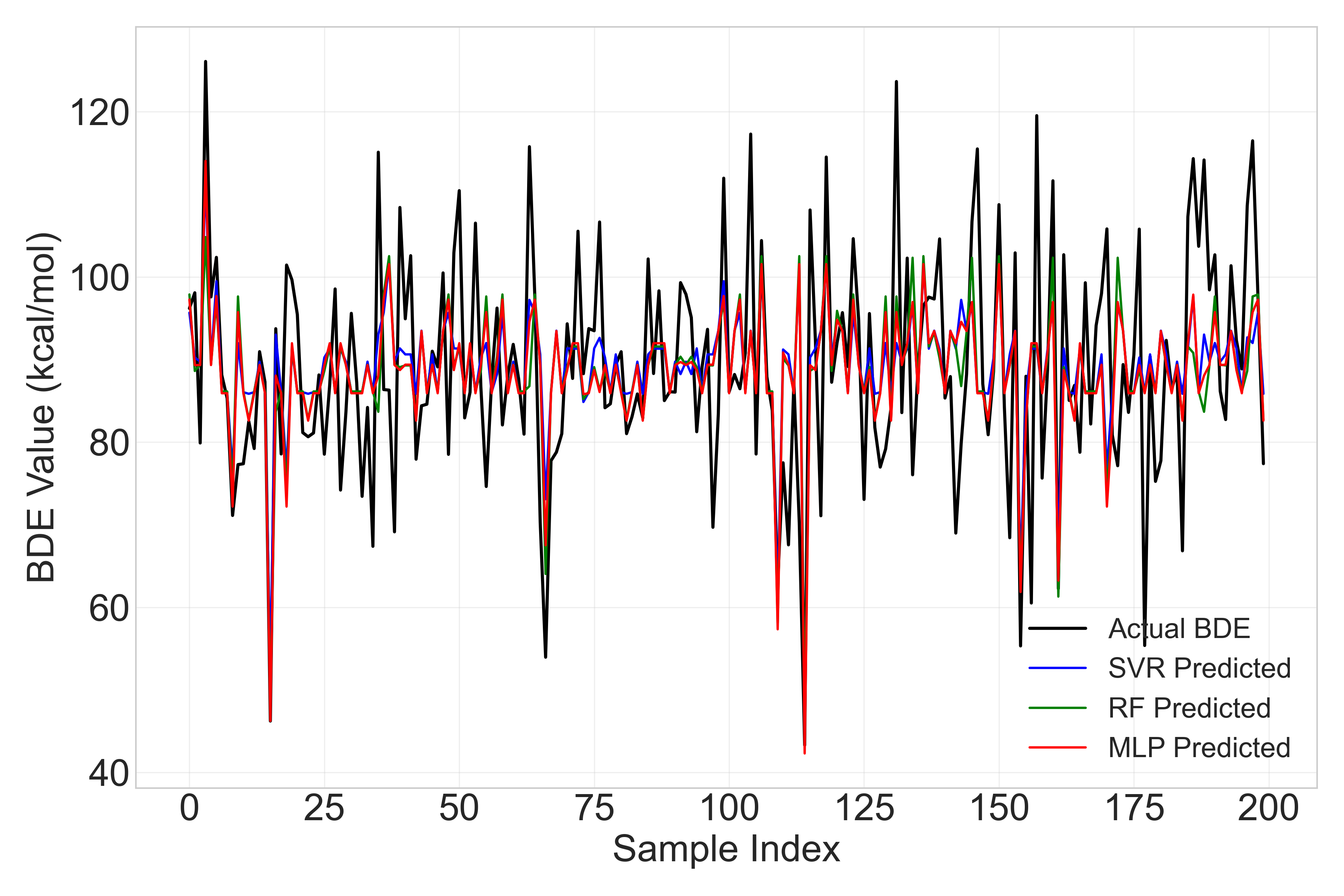}
  \caption{Comparison of Actual vs. Predicted BDE Values Across classical models: SVR, RF and MLP.}
  \label{fig:classicalActualVSPredictedBDE}
\end{figure}

\begin{figure}[h]
  \centering
  \includegraphics[width = \columnwidth]{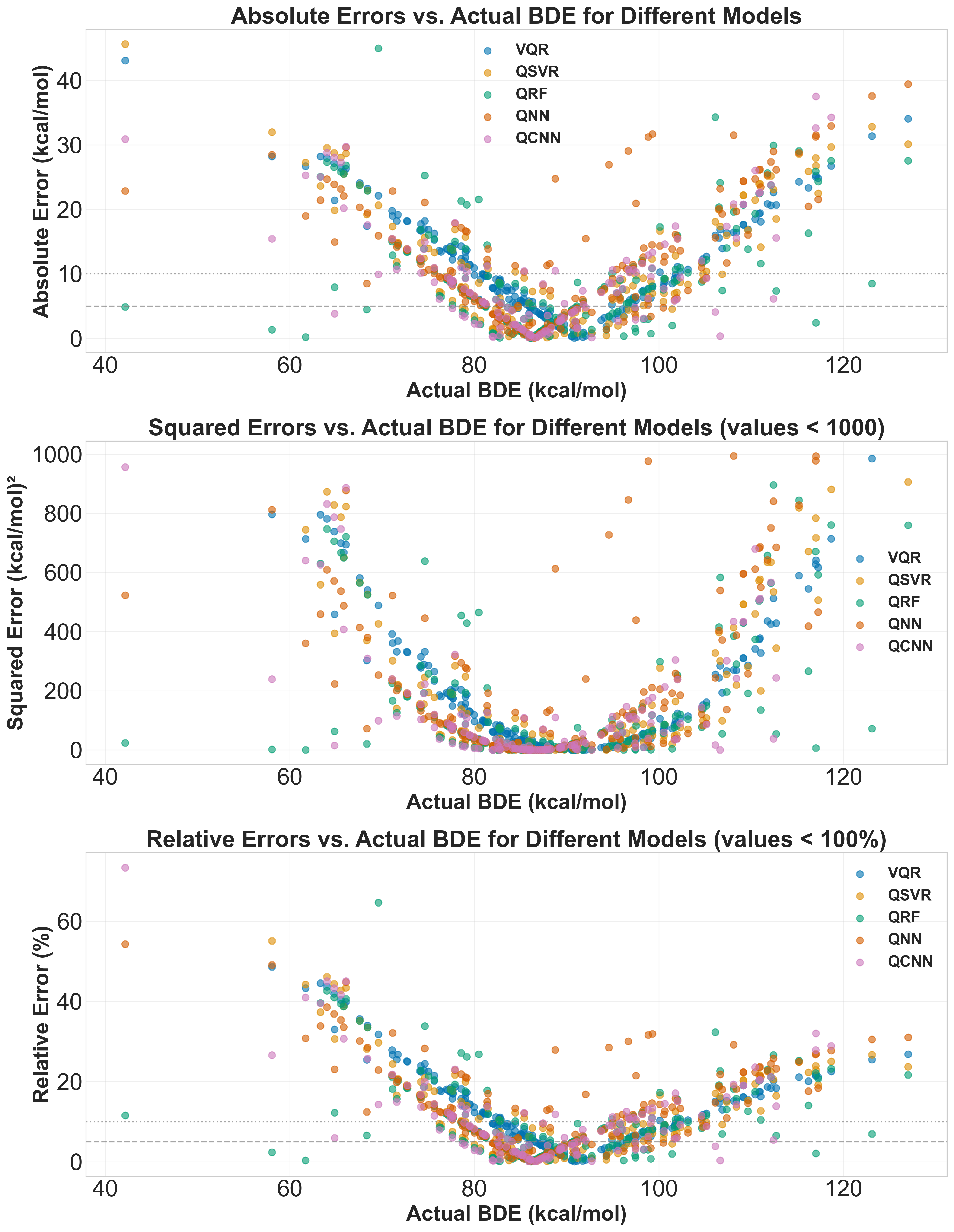}
  \caption{(a) The top panel shows absolute errors (in kcal/mol) versus actual BDE values. All models demonstrate a U-shaped error pattern, with errors minimized around 80-90 kcal/mol and increasing at both lower and higher BDE values. Horizontal reference lines at approximately 5 and 10 kcal/mol indicate error thresholds. (b) The middle panel presents squared errors versus actual BDE, showing a similar U-shaped pattern but with amplified differences at the extremes due to the squared nature of the metric. Values are capped at 1000 (kcal/mol)². (c) The bottom panel displays relative errors (percentage), again exhibiting the U-shaped trend. This visualization highlights that prediction errors are proportionally larger for smaller BDE values, with some errors exceeding 50\% at the lower end of the BDE spectrum.}
  \label{fig:errorComparisonPlots}
\end{figure}

Comparative analysis of predicted BDE values across multiple quantum-inspired models, shown in Fig.s~\ref{fig:ActualVSPredictedBDE}, \ref{fig:classicalActualVSPredictedBDE},  demonstrate strong predictive performance. All five quantum models VQR, QSVR, QRF, QCNN, and QNN capture the general trend of experimental BDEs. In particular, predictions in the mid-range values (approximately 80--95~kcal/mol), which encompass the majority of bonds in the dataset, are consistent and stable. QSVR exhibits minimal fluctuations, indicating robust generalization, while QRF and QCNN are more responsive to local variations, capturing subtle features that other models may overlook. This multi-model quantum approach provides complementary perspectives on bond energetics and highlights the potential of quantum-enhanced predictions. We evaluated the predictive performance of classical (SVR, RF, MLP) and quantum (VQR, QSVR, QRF, QNN, QCNN) models using identical, feature-rich datasets to ensure fair comparison. Performance was assessed through error metrics and threshold-based accuracy measures. Fig.~\ref{fig:errorComparisonPlots} demonstrates the details.

\begin{figure}[h]
  \centering
  \includegraphics[width = \columnwidth]{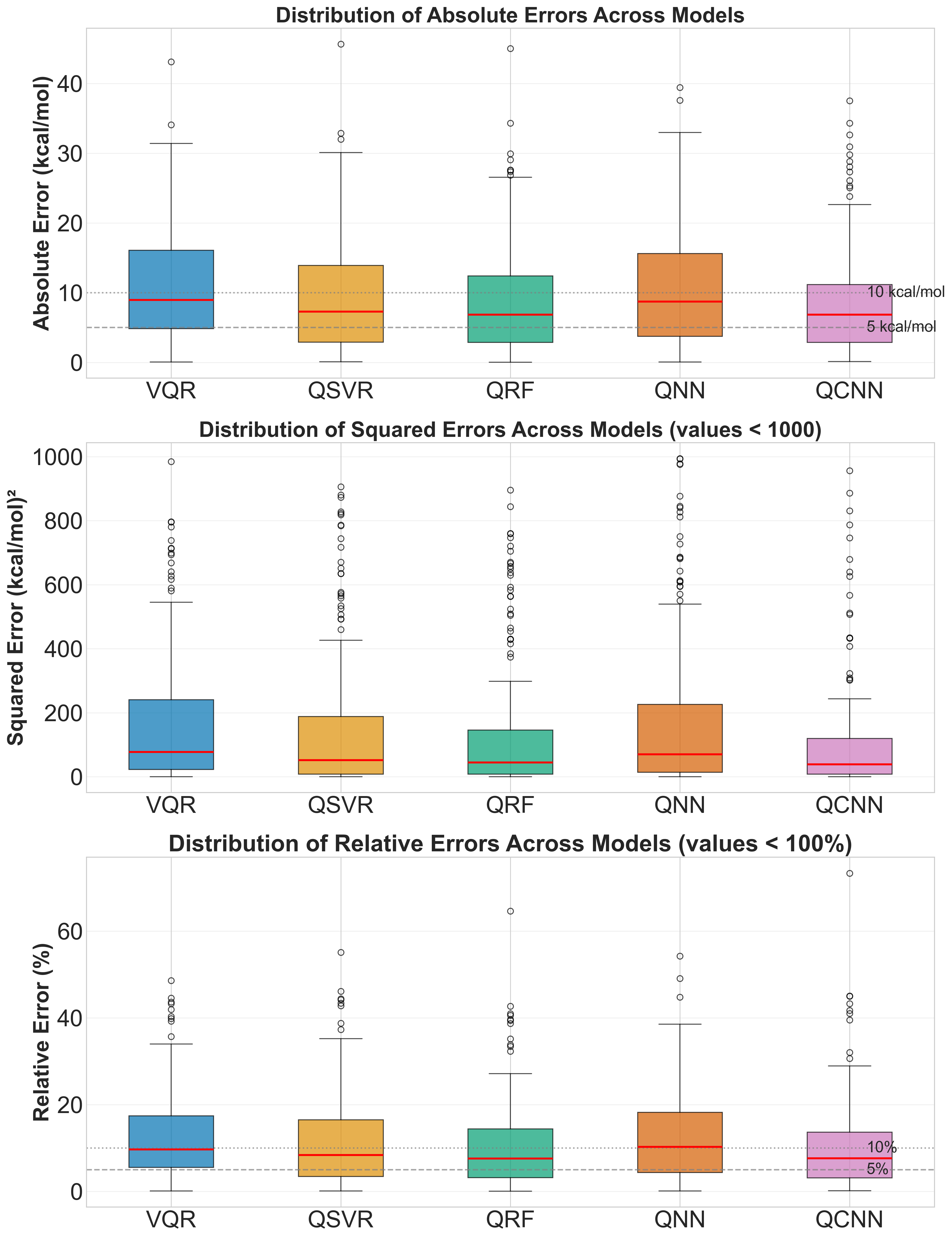}
  \caption{Illustration of the distribution of (top) absolute errors, (middle) squared errors (restricted to values $<$ 1000), 
    and (bottom) relative errors (restricted to values $<$ 100\%) for the VQR, QSVR, QRF, QNN, and QCNN models. 
    Dashed lines represent tolerance thresholds at 5 and 10 kcal/mol (top), as well as at 5\% and 10\% relative error (bottom).}
  \label{fig:errorDistributionPlots}
\end{figure}

\begin{figure}[h]
  \centering
  \includegraphics[width = \columnwidth]{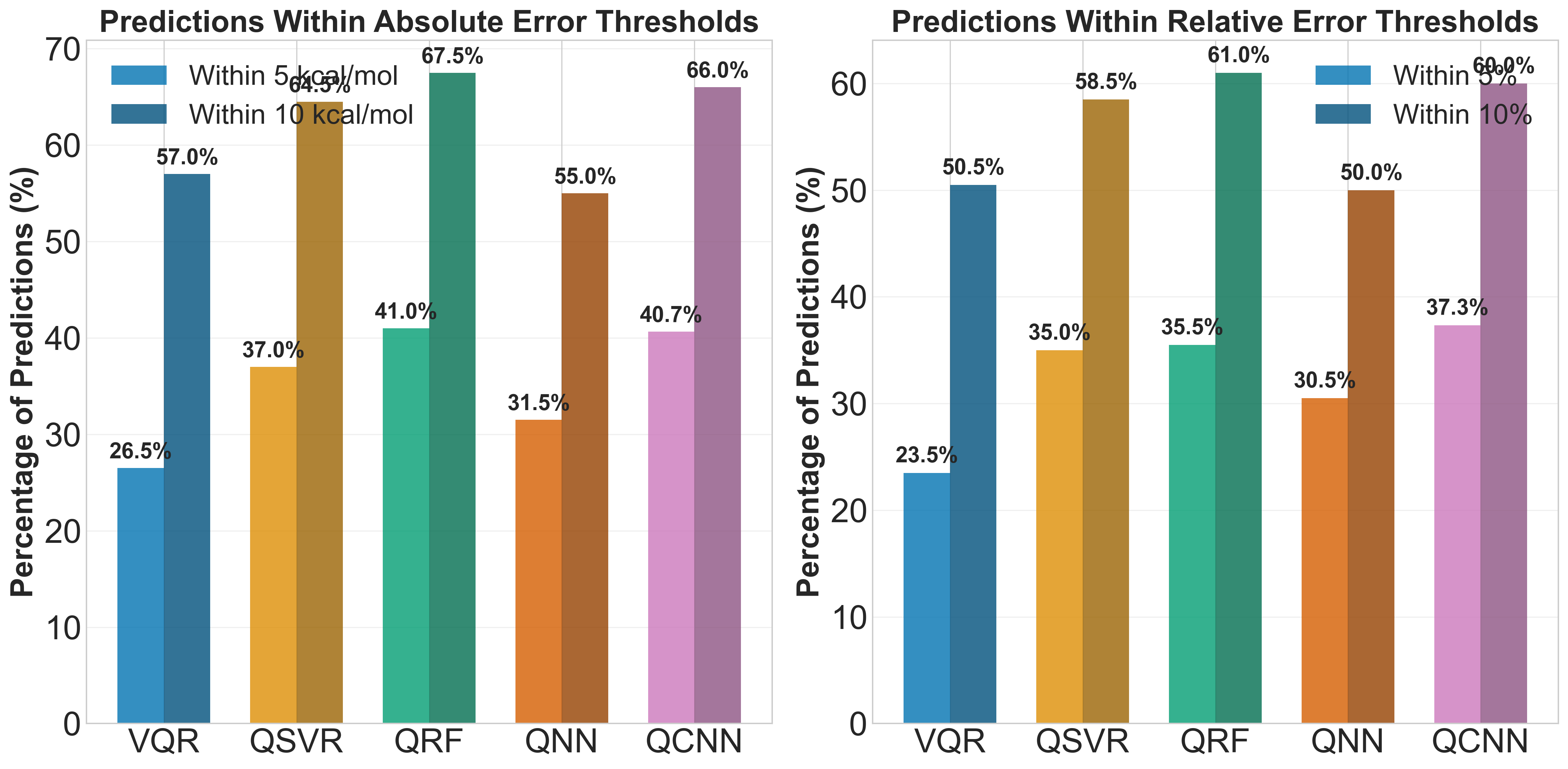}
  \caption{(Left) Percentage of predictions falling within absolute error thresholds of 5 and 10 kcal/mol. 
    (Right) Percentage of predictions falling within relative error thresholds of 5\% and 10\%.}
  \label{fig:errorThresholdPlots}
\end{figure}

Practical utility was evaluated by the proportion of predictions within defined error margins. Table~\ref{tab:performance_metrics} and Fig.s~\ref{fig:errorDistributionPlots},~\ref{fig:errorThresholdPlots} summarize results within absolute thresholds of $\pm 5$~kcal/mol and $\pm 10$~kcal/mol, and relative thresholds of 5\% and 10\%. Among classical models, RF and MLP achieved 41\% and 40\% of predictions within $\pm 5$~kcal/mol, and 67\% and 66\% within $\pm 10$~kcal/mol. Quantum models, notably QRF and QCNN, achieved comparable performance (41\% and 40.7\% within $\pm 5$~kcal/mol; 67.5\% and 66\% within $\pm 10$~kcal/mol), highlighting the efficacy of quantum approaches even in simulation-based environments. Relative error metrics also demonstrate parity between top-performing classical and quantum models, with RF, MLP, QRF, and QCNN achieving 35--37\% of predictions within 5\% error and 60--63\% within 10\% error.

Boxplots of absolute, squared, and relative errors shown in Fig.~\ref{fig:errorDistributionPlots} reveal the robustness of quantum ensemble and convolutional architectures. Median absolute errors for QRF and QCNN remain below 10~kcal/mol, with narrow interquartile ranges indicating consistent performance across diverse molecular structures. Squared error distributions show reduced extreme outliers for QRF and QCNN, mitigating the risk of catastrophic prediction failures. Relative error analysis further confirms that most predictions for these models fall within 10--15\% of the true BDE, emphasizing reliability for practical chemical applications.
  
Error magnitudes vary systematically with BDE value as observed in Fig.~\ref{fig:errorComparisonPlots}. Both classical and quantum models exhibit lowest errors for mid-range BDEs (70--100~kcal/mol), corresponding to well-represented bonds in the training set. Prediction errors increase at the extremes, reflecting data scarcity and structural complexity in these regions, which challenge all classical data-driven approaches.

Table~\ref{tab:performance_metrics} provides detailed quantitative metrics. Classical models SVR, RF, and MLP achieve MSEs of 135.40, 147.88, and 139.92, with corresponding RMSEs of 11.64, 12.16, and 11.83, MAEs near 8.9, and $R^2$ values of 0.22--0.29. Quantum models QCNN and QRF achieve similar MSEs (143.46, 148.72), RMSEs (11.98, 12.19), MAEs (8.85, 9.00), and $R^2$ values (0.14, 0.19), demonstrating that quantum circuits can achieve accuracy comparable to classical ensemble and deep learning methods. Other quantum models, including VQR, QNN, and QSVR, exhibit higher errors and lower $R^2$ values, with VQR showing negative $R^2$, indicative of underfitting.

Observed performance trends correspond closely with the proposed model architecture. Ensemble methods (RF, QRF) aggregate multiple learners, reducing variance and enhancing generalization, resulting in consistently low errors. Deep learning architectures (MLP, QCNN) capture complex nonlinear relationships through hierarchical representations; QCNN leverages quantum convolutional layers to extract features analogous to classical convolutional networks, improving robustness and generalization across diverse bond types. Kernel-based models (SVR, QSVR) effectively project data into higher-dimensional spaces but may underperform relative to ensembles and deep networks for highly nonlinear bond dissociation patterns. Lower performance in VQR and QNN is attributable to challenges in optimizing variational circuits, including barren plateaus and limited expressivity with shallow architectures, which remain active areas of QML research.

\begin{table}[htbp]
\centering
\caption{Performance Metrics Comparison of Classical and Quantum Models}
\begin{tabular}{|p{0.8cm}|p{0.85cm}|p{0.9cm}|p{0.6cm}|p{0.5cm}|p{1.2cm}|p{1.35cm}|}
\hline
\textbf{Model} & \textbf{MSE} & \textbf{RMSE} & \textbf{MAE} & \textbf{R$^2$} & \textbf{5kcal(\%)} & \textbf{10kcal(\%)} \\
\hline
\multicolumn{7}{|c|}{\textit{Classical Models}} \\
\hline
SVR & 135.40 & 11.64 & 8.92 & 0.29 & 40.50 & 65.50 \\
RF & 147.88 & 12.16 & 9.07 & 0.22 & 40.50 & 67.00 \\
MLP & 139.92 & 11.83 & 8.93 & 0.26 & 40.00 & 66.00 \\
\hline
\multicolumn{7}{|c|}{\textit{Quantum Models}} \\
\hline
QCNN & 143.46 & 11.98 & 8.85 & 0.14 & 40.67 & 66.00 \\
QNN & 199.73 & 14.13 & 10.90 & -0.09 & 31.50 & 55.00 \\
QRF & 148.72 & 12.19 & 9.00 & 0.19 & 41.00 & 67.50 \\
QSVR & 167.88 & 12.96 & 9.71 & 0.08 & 37.00 & 64.50 \\
VQR & 176.71 & 13.29 & 10.76 & 0.03 & 26.50 & 57.00 \\
\hline
\end{tabular}
\label{tab:performance_metrics}
\end{table}

\section{Towards near-chemical accuracy}

We have the following insightful observation towards improving accuracy in the BDE prediction. Achieving MAE on the order of around \textit{1 kcal/mol} for BDE prediction requires a carefully designed pipeline that integrates high-fidelity reference data, chemically expressive representations, robust learning architectures, and rigorous evaluation. In the following, we outline a scientifically grounded strategy for attaining such accuracy.

The fidelity of the training labels bounds the predictive performance of any supervised model. To approach 1 kcal/mol MAE, label noise must be comparably small. We recommend augmenting or replacing density functional theory (DFT) labels with higher-level quantum chemical references for a curated ``gold standard'' subset, using methods such as DLPNO-CCSD(T), CCSD(T)-F12, or composite protocols (e.g., W1--W4, G4(MP2)-6X). All thermochemical quantities should be computed with consistent protocols, including geometry optimizations, harmonic frequency calculations for zero-point energy (ZPE), and thermal corrections to enthalpy at 298\,K. BDEs should be computed consistently as
\[
\Delta H_\text{BDE} = H(\mathrm{radical}\,A) + H(\mathrm{radical}\,B) - H(\mathrm{parent}),
\]
with properly treated spin states and identical levels of theory for parent and radicals. Where available, experimental benchmarks should be incorporated to perform post hoc calibration (e.g., isotonic regression or shallow residual models), thereby aligning computational predictions with empirical scales.

BDEs are sensitive to subtle electronic and environmental effects, including induction, resonance, conjugation, and ring strain. To reduce irreducible error, representations should incorporate:
\begin{enumerate}
    \item \textit{Physics-informed descriptors:} local orbital populations (e.g., NBO charges), bond order metrics (e.g., Wiberg/Mayer indices), local spin densities on fragments, frontier orbital energies, and electrostatic potentials at nuclei proximal to the bond.
    \item \textit{Local environment features:} ring membership and strain indicators, aromaticity and conjugation flags, and descriptors that partition $\sigma$ and $\pi$ contributions near the dissociating bond.
    \item \textit{Geometric context:} 3D features such as bond lengths, bond angles, dihedrals, pyramidalization, and Boltzmann-weighted averages across low-energy conformers.
    \item \textit{Learned molecular embeddings:} message passing neural networks (MPNNs), rotationally and translationally equivariant graph neural networks (e.g., PaiNN, NequIP, Allegro), or transformer-based architectures trained directly on 3D molecular graphs with bond-centric pooling.
\end{enumerate}
These features capture both local and nonlocal effects pertinent to bond strength and are critical for improving accuracy.

To minimize both bias and variance, strong classical baselines should be employed alongside advanced deep learning models:
\begin{enumerate}
    \item \textit{Classical learners:} gradient boosting machines (e.g., XGBoost, LightGBM, CatBoost) with careful regularization provide robust nonparametric baselines.
    \item \textit{Deep architectures:} MPNNs and equivariant GNNs trained on 3D structures can capture complex, non-linear structure--property relationships. Incorporating uncertainty (e.g., through ensembling or Monte Carlo dropout) improves reliability.
    \item \textit{Multi-fidelity learning:} a large dataset at a consistent DFT level is paired with a smaller, high-fidelity set at the CCSD(T)-quality level. The model learns corrections via $\Delta$-learning: $\Delta \text{BDE} = \text{BDE}_{\text{high}} - \text{BDE}_{\text{DFT}},$ which are added to fast DFT-level predictions at inference time.
    \item \textit{Stacking and ensembling:} combining physically motivated linear/bond-order models with non-linear learners (GNNs/GBDTs) reduces variance and corrects systematic biases. Ensembles of independently initialized neural networks further tighten error distributions.
\end{enumerate}

Sustained accuracy near 1 kcal/mol is achievable in-domain; coverage gaps typically drive large errors. Training sets should be diversified to span bond classes, substitution patterns, and electronic environments. Diversity sampling (e.g., farthest point sampling in learned embedding spaces) helps maintain coverage. Active learning loops can be employed to identify high-uncertainty cases for which high-level labels (e.g., DLPNO-CCSD(T)) are computed and incorporated to refine the model. Data augmentation strategies (e.g., conformer perturbations, small chemically valid edits) combined with fast $\Delta$-learning corrections may further improve generalization when label fidelity is preserved. Evaluation must prevent leakage and over-optimistic estimates. Scaffold- or bond-class-based splits are recommended, with per-class metrics reported for C--H, C--C, C--O, and other bond types. Both absolute (kcal/mol) and relative (\%) errors should be monitored, with particular attention to extreme BDE regimes (e.g., $<60$ and $>110$ kcal/mol). Calibrated uncertainty quantification (e.g., conformal prediction or deep ensembles) should be provided to flag high-risk predictions for backup with ab initio calculations.

A realistic pathway comprises:
\begin{enumerate}
    \item \textit{Curate a multi-fidelity corpus:} gather $5\times 10^4$--$2\times 10^5$ bonds at a consistent DFT level (e.g., $\omega$B97X-D/def2-TZVP with ZPE/thermal corrections), and 2\,000--10\,000 bonds at a DLPNO-CCSD(T)/CBS or CCSD(T)-F12 level for calibration.
    \item \textit{Architect a strong learner:} train an equivariant GNN on 3D conformers augmented with bond-centric quantum descriptors; optionally include auxiliary tasks (e.g., fragment atomization energies, spin densities) to regularize representations.
    \item \textit{$\Delta$-learning:} learn the high-level correction $\Delta \text{BDE}$ on the gold set and add it to DFT predictions at inference.
    \item \textit{Ensembling and calibration:} average predictions from 5--10 independently trained GNNs; apply isotonic regression or similar calibration against the high-fidelity set.
    \item \textit{Active learning:} iteratively acquire high-level labels for high-uncertainty bonds until validation MAE plateaus near $\sim$1 kcal/mol across bond classes.
\end{enumerate}

Although current NISQ-era hardware imposes practical constraints, quantum components can provide complementary benefits:
\begin{enumerate}
    \item \textit{Quantum kernels:} integrate quantum kernels based on problem-tailored feature maps into multiple-kernel learning alongside classical kernels to capture subtle correlation patterns.
    \item \textit{Variational $\Delta$-correctors:} train shallow variational circuits (e.g., VQR or QCNN) to learn $\Delta$-corrections between DFT and high-level methods for bond-centric fragments, using robust optimizers and parameter regularization to mitigate barren plateaus.
    \item \textit{Quantum-derived descriptors:} compute small active-space fragment properties (e.g., entanglement measures or two-particle RDM features) and feed them into classical regressors.
\end{enumerate}
In practice, the dominant accuracy gains will stem from improved labels and representations, with quantum components serving as complementary augmentations under resource constraints. Key obstacles include label inconsistency and protocol mismatch, conformational sensitivity, data imbalance for rare bonds, and overfitting in deep models. These can be mitigated via automated and consistent quantum-chemistry workflows, Boltzmann-weighted conformer features applied consistently to parent and radical species, stratified sampling with targeted active learning for rare chemistries, and robust regularization (early stopping, weight decay, dropout) under scaffold-aware splits.

The primary determinants of achieving $\sim$1 kcal/mol MAE are (i) high-fidelity and consistent reference data, (ii) physics-informed and geometrically expressive representations, (iii) multi-fidelity $\Delta$-learning with ensembling, and (iv) coverage-aware training enhanced by active learning and uncertainty quantification. Within these guardrails, sustained near-chemical accuracy is feasible in-domain, while rigorous validation and calibrated uncertainties enable safe deployment across more diverse chemical spaces.
\section{Concluding Remarks}
We presented a unified, reproducible benchmark comparing classical and quantum machine learning for bond dissociation energy prediction, demonstrating that quantum models especially QCNN and QRF achieve accuracy and robustness comparable to strong classical baselines (RF, MLP), particularly in the mid-range BDE regime. Error analyses reveal systematic challenges at distribution extremes, underscoring the importance of coverage, chemically expressive representations, and rigorous evaluation. Our findings motivate a targeted path to near-chemical accuracy: curating high-fidelity labels (e.g., DLPNO-CCSD(T)/CBS), enriching bond-centric and 3D features, adopting multi-fidelity $\Delta$-learning with calibrated ensembles, and leveraging active learning to close domain gaps. Within this framework, quantum components serve as complementary augmentations (quantum kernels, variational $\Delta$-correctors, quantum-derived descriptors), offering useful inductive biases under current constraints. Overall, this work establishes a practical baseline for quantum-enhanced molecular property prediction and outlines concrete priorities to approach $\sim$1 kcal/mol MAE in BDE prediction.

\bibliographystyle{ieeetr}
\bibliography{ref}


\end{document}